\documentstyle[twoside,fleqn,espcrc2]{article}

\newcommand{\AmS}{{\protect\the\textfont2
  A\kern-.1667em\lower.5ex\hbox{M}\kern-.125emS}}

\title{Staggered fermions and their $O(a)$ improvements}

\author{Yubing Luo\address{Columbia University, Department of
	Physics, New York, NY 10027} 
	\hfill \raisebox{4.0cm}[0cm][0cm]{CU-TP-779} }

\begin{document}

\begin{abstract}
Expanding upon the arguments of Sharpe, we explicitly implement the
Symanzik improvement program demonstrating the absence of order $a$
terms in the staggered fermion action. We propose a general program
to improve fermion operators to remove $O(a)$ corrections from their
matrix elements, and demonstrate this program for the examples of
matrix elements of fermion bilinears and $B_K$. We also determine
the additional operators which must be added to improve the standard
staggered fermion currents.
\end{abstract}

\maketitle

\section{INTRODUCTION}

In order to reduce the systematic errors coming from finite
lattice spacing, we must improve both the action and the lattice
operators. According to the improvement program of Symanzik, 
we find all dimension-5 operators which are invariant under
the lattice symmetry group and add them to the original action
to reduce all $O(a)$ corrections. If there exists no such 
dimension-5 operator, then the action is already good to
$O(a^2)$. On the other hand, even though the action is accurate
to $O(a^2)$, the matrix elements generally may still have
$O(a)$ errors, and hence, we have to improve the
operators themselves.

In this paper, I will expand upon the arguments given by 
Sharpe\cite{Sharpe_1} to prove that there are no $O(a)$ 
terms which can be
added to the staggered fermion action, and hence the action is
already accurate to $O(a^2)$. Then I will propose a set of improved
fermion field variables and use them to construct the fermion
operators to reduce the $O(a)$ corrections to their matrix elements.
We apply this program to the case of 
$\langle 0|\bar{s}\gamma_{54}d|K^0 \rangle$
and $B_K$ as examples. We find the former does have $O(a)$
corrections while the latter does not.
We will also determine the additional operators that must be added
to improve the standard staggered fermion currents to define
operators whose matrix elements are accurate to $O(a^2)$.

\section{STAGGERED FERMION ACTION IS ACCURATE TO $O(a^2)$}

What we will prove here is that there exists no dimension-5 operator
which is invariant under the lattice symmetry group (rotations, axis
reversal, translation, $U(1) \otimes U(1)$, charge conjugation,
etc), and therefore, the staggered fermion action is already good to
$O(a^2)$.

We rewrite the staggered fermion action as
\begin{equation}
S_f = \bar{\chi} [\sum_\mu \overline{(\gamma_\mu \otimes I)} 
	D_\mu + m ] \chi,
\end{equation}
and consider all dimension-5 operators which have the general form
$\bar{\chi} \overline{\gamma_S \otimes \xi_F} f(D)
\chi$ where $f$ is a homogenous real polynomial of degree 2.

Invariance under $U_A(1)$ requires that $S+F$ is odd, so only the
following combinations of $S \otimes F$ are valid:
\begin{equation}
	(I, \gamma_5, \gamma_{\mu\nu}, \gamma_{5\mu\nu}) \otimes 
	(\xi_\lambda, \xi_{5\lambda}),
\end{equation}
\begin{equation}
	(\gamma_\mu, \gamma_{5\mu}) \otimes 
	(I, \xi_5, \xi_{\lambda\tau}, \xi_{5\lambda\tau}).
\end{equation}

The axis reversal invariance will further limit the number of terms
to be four:
\begin{equation}\label{term:1}
   \gamma_5 \otimes \xi_\mu D^2_\mu,
\end{equation}
\begin{equation}\label{term:2}
   \gamma_5 \otimes \xi_{5\mu} D^2_\mu,
\end{equation}
\begin{equation}\label{term:3}
   \gamma_5[\gamma_\mu, \gamma_\nu] \otimes \xi_5(\xi_\mu +
    	\xi_\nu) [D_\mu, D_\nu],
\end{equation}
\begin{equation}\label{term:4}
   \gamma_5[\gamma_\mu, \gamma_\nu] \otimes \xi_5(\xi_\mu -
    	\xi_\nu) \{D_\mu, D_\nu\}.
\end{equation}

The rotational invariance will further eliminate the term in
Eq.(\ref{term:1}) but allows the remaining three terms
Eq.(\ref{term:2} - \ref{term:4}).

Finally, none of the terms listed in
Eq.(\ref{term:2} - \ref{term:4}) are invariant under 
lattice translation! So, we conclude that there is no dimension-5 
fermion operator which is invariant under the lattice symmetry group,
and therefore no dimension-5 operator can be added to the
staggered fermion action.

\section{IMPROVE STAGGERED FERMION FIELDS}

Following ref.\cite{kluberg-stern}, we define the hypercubic 
fields as
\begin{equation} {\label{cov_field_1}}
    q(y) = \frac{1}{2}\sum_A\gamma_A{\cal{U}}_A(y)\chi_A(y),
\end{equation}
\begin{equation} {\label{cov_field_2}}
    \bar{q}(y) = \frac{1}{2}\sum_A\bar{\chi}_A(y){\cal{U}}^\dagger_A(y)
    	\gamma_A^\dagger.
\end{equation}
Then the classical continuum limit of the staggered fermion action
can be written as:
\begin{eqnarray}
    S_F[q, \bar{q}] \rightarrow \int_y \sum_{AB} \bar{q}(y) 
    	\{ \sum_{\mu} (\gamma_\mu \otimes I)_{AB} D_\mu 
    		\nonumber \\
    + m\delta_{AB} -a [ \sum_\mu (\gamma_5 \otimes 
    	\xi_{5\mu})_{AB}D^2_\mu
    		\nonumber \\
    +\frac{ig_0}{4} \sum_{\mu\nu}((\gamma_\mu-\gamma_\nu) \otimes I 
    + \frac{1}{2}\gamma_5 [\gamma_\mu, \gamma_\nu]
    		\nonumber \\
    \otimes \xi_5(\xi_\mu + \xi_\nu))_{AB}F_{\mu\nu}(y)]\} q(y)
    	\nonumber \\
    +O(a^2).	\qquad \qquad \qquad \qquad \qquad \qquad 
\end{eqnarray}
where $D_\mu=\partial_\mu+igA_\mu$ is the continuum covariant
derivative. At first sight, the action contains order $a$ terms.
Likewise, it is clear that the fermion propagator for the 
hypercubic fields $q$ and $\bar{q}$ deviates from the
continuum propagator by terms of order $a$.
However, if we introduce the following improved field variables
\begin{equation}
    \chi^I_A(y) = (1-a\sum_\nu A_\nu D^L_\nu)\chi_A(y),
\end{equation}
\begin{equation}
    \bar{\chi}^I_A(y)=\bar{\chi}_A(y)(1-a\sum_\nu A_\nu 
    	\stackrel{\scriptstyle{\leftarrow}
    \scriptstyle{L}} {D_\nu}),
\end{equation}
and replace $\chi$, $\bar{\chi}$ in Eq.(\ref{cov_field_1}, 
\ref{cov_field_2}) by 
$\chi^I$ and $\bar{\chi}^I$, we will reduce the finite lattice
spacing corrections from $O(a)$ to $O(a^2)$.

Using the new fermion fields, we can construct improved fermion
operators. For example, the improved fermion bilinears have the
following form:
\begin{eqnarray} \label{improved_bilinear}
  \bar{\chi}^I_A(y) \overline{(\gamma_S 
     \otimes \xi_F)}_{AB} \chi^I_B(y) = \qquad \qquad \qquad
     	\nonumber       \\
  \bar{\chi}_A(y)\overline{(\gamma_S \otimes \xi_F)}_{AB} 
     \chi_B(y)		\nonumber	\\
  - \frac{a}{2}\sum_\nu \partial^L_\nu [\bar{\chi}_A(y)
     (\overline{(\gamma_S \otimes \xi_F)}_{AB} 
     	\nonumber       \\
  - \overline{(\gamma_{5\nu S} \otimes \xi_{5\nu F})}_{AB})
        \chi_B(y)]	\nonumber	\\
  - \frac{a}{2}\sum_\nu \bar{\chi}_A(y)
       [\overline{(\gamma_{5\nu S} \otimes \xi_{5\nu F})}_{AB}
     	\nonumber       \\
  - \overline{(\gamma_{S5\nu} \otimes \xi_{F5\nu})}_{AB}]
       D^L_\nu \chi_B(y)	\nonumber	\\
  +O(a^2). \qquad \qquad \qquad \qquad \qquad 
\end{eqnarray}

\section{APPLICATIONS}

Here, we apply this improvement program to the calculation of 
matrix element 
$\langle 0|\bar{s}\gamma_{54}d|K^0 \rangle$, 
the calculation of $B_K$, and the improvement of lattice currents.

\subsection{$\langle 0|\bar{s}\gamma_{54}d|K^0 \rangle$} 
\label{application:1}
The axial current used in the (Landau gauge) numerical simulation is:
\begin{equation}
    A_\mu(y) = \sum_{AB}\bar{\chi}_A \overline{(\gamma_{5\mu} \otimes
    \xi_5)}_{AB} \chi_B(y).
\end{equation}
From the continuum expression
\begin{eqnarray}
    P(t)^{cont} = \langle 0|A_4(t)^{cont}|K^0 \rangle \nonumber \\
     	= \sqrt{2}f_K m_k e^{-m_k|t|},
\end{eqnarray}
we define, on the lattice,
\begin{equation}
    P(t) = \langle 0|\sum_{\vec{x}}A_4(\vec{x},t)|K^0 \rangle ,
\end{equation}
and put the wall source that creates the $K^0$ on the time 
slice at $t=0$. If we don't consider $O(g_0^2 a)$ terms, we can take
only the term $-\frac{a}{2}\sum_\nu \partial_\nu^L [\bar{\chi}_A
\overline{(\gamma_{54} \otimes \xi_5)}_{AB}\chi_B]$ in
Eq.(\ref{improved_bilinear}) because other terms contribute zero
``flavor'' trace at the tree level. So, we have
\begin{equation}
    P(t)^{Imp}  =  P(t) - \frac{a}{2}\partial_4^L P(t),
\end{equation}
and this simple example show that, if we do not consider $O(g_0^2a)$
corrections, the improved operator gives the well-known
$t+\frac{1}{2}$ interpretation:
\begin{equation}
    P(t) = \sqrt{2}f_K m_k e^{-m_k|t+1/2|}.
\end{equation}

\subsection{$B_K$}
The formula for calculating $B_K$ is:
\begin{equation}
    B_K = \frac{{\cal{M}}_K}{\frac{8}{3} {\cal{M}}_K^V}.
\end{equation}
The improved numerator is (omitting the $O(g_0^2a)$ terms):
\begin{equation}
    {\cal{M}}_K^{Imp} = {\cal{M}}_K - \frac{a}{2} \partial_4^L
        {\cal{M}}_K + O(a^2).
\end{equation}
Since ${\cal{M}}_K(t)$ is computed from a plateau 
(i.e. time independent) within the statistical error,
there is no $O(a)$ corrections to the numerator.
If we calculate the quantity $\langle 0|\bar{s}\gamma_4\gamma_5 
d|K^0 \rangle $ in
both forward and backward directions and multiply them to get the
denominator, then there will have no $O(a)$ correction even if no
attention is paid to an accurate definition of $f_K$.
Hence, we showed that there is neither $O(a)$ 
nor $O(ag_0^{2n}\log^na)$ corrections to $B_K$.
Sharpe \cite{Sharpe_1} has examined this question in greater detail
and argued that in fact there are no corrections of $O(g_0^{2n}a)$
also.  However, if we calculated the denominator 
only in one time direction, there would be an error of
order of $O(m_ka)$.

\subsection{Renormalization of lattice currents}
The lattice currents can be written as\cite{currents}:
\begin{equation}
    J^F_{latt} = \bar{\chi} \overline{\gamma_J \otimes
    		\xi_F} \chi,
\end{equation}
Using the method developed in this paper, we can explicitly
determine the improved currents accurate to $O(a^2)$. For example,
the conserved vector current and axial vector current corresponding
to the $U_V(1) \otimes U_A(1)$ lattice symmetry can be written as
follows:
\begin{eqnarray}
    V_\mu^I(y) = V_\mu(y) -\frac{a}{2}\sum_\nu \partial_\nu 
    		[\bar{\chi}_A (\overline{\gamma_\mu \otimes
    		I})_{AB} \chi_B]  \nonumber \\
    		-\frac{a}{2}\partial_\mu 
    		[\bar{\chi}_A (\overline{\gamma_\mu \otimes
    		I})_{AB} \chi_B]  \nonumber \\ 
    		-\frac{a}{4}\sum_\nu \partial_\nu 
    		[\bar{\chi}_A (\overline{\gamma_{5[\mu,\nu]} 
    		\otimes \xi_{5\nu}})_{AB} \chi_B] 
    			\nonumber \\
    		+ O(a^2), \qquad \qquad \qquad \qquad \qquad
    		\qquad
\end{eqnarray}
\begin{eqnarray}
    A_\mu^{\xi_5,I}(y) = A_\mu^{\xi_5}(y) 
    	\qquad \qquad \qquad \qquad \qquad \qquad \nonumber \\
    		-\frac{a}{2}\sum_\nu \partial_\nu 
    		[\bar{\chi}_A (\overline{\gamma_{\mu 5} \otimes
    		\xi_5})_{AB} \chi_B]  \nonumber \\
    		-\frac{a}{2}\partial_\mu 
    		[\bar{\chi}_A (\overline{\gamma_{\mu 5} \otimes
    		\xi_5})_{AB} \chi_B]  \nonumber \\
    		+\frac{a}{4}\sum_\nu \partial_\nu 
    		[\bar{\chi}_A (\overline{\gamma_{[\mu,\nu]} 
    		\otimes \xi_{\nu}})_{AB} \chi_B]
    			\nonumber \\
    		+ O(a^2). \qquad \qquad \qquad \qquad \qquad \qquad
\end{eqnarray}
The effect of the second term on the right hand side is to shift the
position y, labeling the current, from the corner to the center of
the hypercube. The third term whose effect is to shift in the
$\mu$'s direction occurs here because the currents are non-local
operators which involve an overlap between two nearest
hypercubes. The forth term is a mixing of a different spin-flavor
operator and is necessary to remove all order $a$ effects from a
general matrix element.

\section{CONCLUSIONS}

The staggered fermion action is accurate to $O(a^2)$;
In order to reduce $O(a)$ corrections in a matrix element, we
must use the improved fermion fields which we have proposed;
Our improved program derived the well-known
$t+\frac{1}{2}$ interpretation for the one hypercubic fermion
bilinears at the tree-level; We demonstrated
$B_K$ is good to $O(g_0^2a)$ at the tree-level; 
We proposed terms which must to included in
the lattice currents in order to reduce $O(a)$ corrections.

\bigskip
\bigskip

I warmly thank Prof. Norman H. Christ for the
extensive discussions during every stage of this work.
I am also grateful to Weonjong Lee for the numerical data. 
I also thank Bob Mawhinney for the helpful discussions
on the lattice symmetry of staggered fermions.

\end{document}